\documentclass[sigchi, authorversion=true, nonacm=true]{acmart}

\AtBeginDocument{%
  \providecommand\BibTeX{{%
    \normalfont B\kern-0.5em{\scshape i\kern-0.25em b}\kern-0.8em\TeX}}}

\setcopyright{acmcopyright}
\copyrightyear{2021}
\acmYear{2021}
\acmDOI{10.1145/1122445.1122456}

\acmConference[ISWC '21]{ISWC '21: International Symposium on Wearable Computers}{Sept, 2021}{}
\acmBooktitle{ISWC '21: International Symposium on Wearable Computers, Sept, 2021}
\acmPrice{15.00}
\acmISBN{978-1-4503-XXXX-X/18/06}
\usepackage{subcaption}

\begin{document}

\title{Improving Deep Learning for HAR with shallow LSTMs}

\author{Marius Bock}
\affiliation{%
  \institution{University of Siegen}
  \streetaddress{H\"olderlinstr. 3}
  \city{Siegen}
  \state{North Rhine-Westphalia}
  \postcode{57076}
  \country{Germany}
}
\email{marius.bock@uni-siegen.de}

\author{Alexander H\"olzemann}
\affiliation{%
  \institution{University of Siegen}
  \streetaddress{H\"olderlinstr. 3}
  \city{Siegen}
  \state{North Rhine-Westphalia}
  \postcode{57076}
  \country{Germany}
}
\email{alexander.hoelzemann@uni-siegen.de}

\author{Michael Moeller}
\affiliation{%
  \institution{University of Siegen}
  \streetaddress{H\"olderlinstr. 3}
  \city{Siegen}
  \state{North Rhine-Westphalia}
  \postcode{57076}
  \country{Germany}
}
\email{michael.moeller@uni-siegen.de}

\author{Kristof Van Laerhoven}
\orcid{0000-0001-5296-5347}
\affiliation{%
  \institution{University of Siegen}
  \streetaddress{H\"olderlinstr. 3}
  \city{Siegen}
  \state{North Rhine-Westphalia}
  \postcode{57076}
  \country{Germany}
}
\email{kvl@eti.uni-siegen.de}


\begin{abstract}
Recent studies in Human Activity Recognition (HAR) have shown that Deep Learning methods are able to outperform classical Machine Learning algorithms. One popular Deep Learning architecture in HAR is the DeepConvLSTM. In this paper we propose to alter the DeepConvLSTM architecture to employ a 1-layered instead of a 2-layered LSTM. We validate our architecture change on 5 publicly available HAR datasets by comparing the predictive performance with and without the change employing varying hidden units within the LSTM layer(s). Results show that across all datasets, our architecture consistently improves on the original one: Recognition performance increases up to $11.7\%$ for the F1-score, 
and our architecture significantly decreases the amount of learnable parameters. This improvement over DeepConvLSTM decreases training time by as much as $48\%$. Our results stand in contrast to the belief that one needs at least a 2-layered LSTM when dealing with sequential data. Based on our results we argue that said claim might not be applicable to sensor-based HAR. 
\end{abstract}



\maketitle

\section{Introduction}
Physical activities play a crucial role in the way we structure our lives. Which activity, and how it is performed, can reveal a person's intention, habit, fitness, and state of mind; it is therefore not surprising that a range of research fields, from cognitive science to healthcare, display a growing interest in the machine recognition of human activities, also known as Human Activity Recognition (HAR) \cite{bullingTutorialHumanActivity2014}.
Deep Learning methods have in the past decade shown to outperform classical Machine Learning algorithms (e.g., \cite{krizhevskyImagenetClassificationDeep2012, collobertNaturalLanguageProcessing2011, szegedyGoingDeeperConvolutions2015}) and, as a product of this success, have led to studies investigating the effectiveness of Deep Learning in HAR (e.g., \cite{ordonezDeepConvolutionalLSTM2016, hammerlaDeepConvolutionalRecurrent2016}). 

One of the most popular Deep Learning architectures for HAR is the DeepConvLSTM which was proposed by Ordonez et al. in \cite{ordonezDeepConvolutionalLSTM2016}. The suggested architecture combines both recurrent and convolutional layers and received state-of-the-art results on both the Opportunity \cite{roggenCollectingComplexActivity2010} and the Skoda Mini Checkpoint \cite{zappiActivityRecognitionOnbody2008} dataset. The original DeepConvLSTM architecture employs a 2-layered LSTM with $128$ hidden units. A common belief, as e.g. stated by Chen et al. \cite{chenDeepLearningSensorbased2021}, based on the findings of Karpathy et al. \cite{karpathyVisualizingUnderstandingRecurrent2015}, is that one requires at least a 2-layered LSTM when dealing with sequential data. With this paper, we aim at challenging this belief and suggest that re-examining the architecture of the DeepConvLSTM by employing a one-layered LSTM, has considerable benefits.

Our paper's contributions are threefold:
\begin{enumerate}
    \item We show that an altered DeepConvLSTM architecture with a one-layered LSTM overall outperforms architectures employing a two-layered LSTM, by validating our claim using the Opportunity dataset \cite{roggenCollectingComplexActivity2010} as seen in \cite{ordonezDeepConvolutionalLSTM2016}, as well as 4 other popular HAR datasets \cite{schollWearablesWetLab2015, sztylerOnBodyLocalizationWearable2016, reyes-ortizTransitionAwareHumanActivity2016, stisenSmartDevicesAre2015}.
    \item Using our suggested architecture change, we reduce the number of learnable parameters within the DeepConvLSTM and thus are able to decrease training time significantly.
    \item We provide our PyTorch-based architecture, experiment scripts and log files in a GitHub repository for others to replicate these findings and continue such analyses.
\end{enumerate}

\begin{figure*}[h]
  \centering
  \includegraphics[width = 0.99\linewidth]{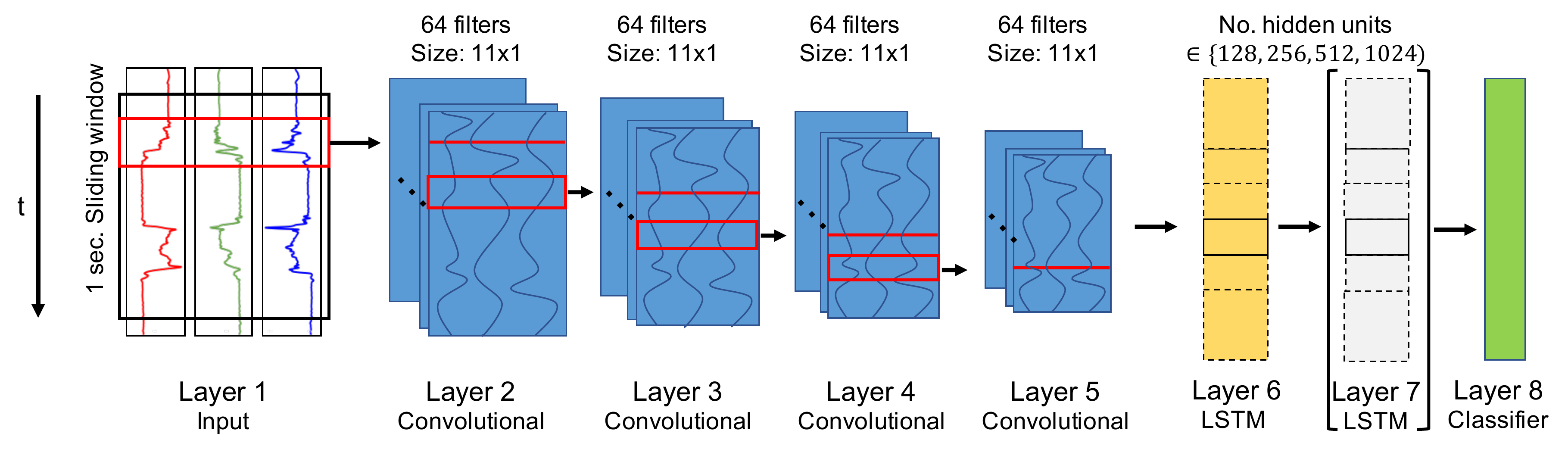}
  \caption{Illustration of the suggested change to the DeepConvLSTM  \cite{ordonezDeepConvolutionalLSTM2016} architecture. The change involves removing the second LSTM layer (Layer 7). During experiments we either include (original architecture) or exclude Layer 7 (our suggestion) and vary the amount of hidden units within the LSTM layer(s) (i.e. Layers 6 and 7) to be either 128, 256, 512 or 1024.}
  \Description{Illustration of the suggested change to the DeepConvLSTM architecture, i.e. removing the second LSTM layer, and varying hidden units within experiments.}
  \label{fig:architecture}
\end{figure*}

\section{Related Work}
\emph{Deep Learning in HAR.}
The predictive performance of classical Machine Learning approaches highly relies on sophisticated, handcrafted features \cite{pouyanfarSurveyDeepLearning2018}. In the last decade, Deep Learning has shown to outperform classical Machine Learning algorithms in many areas, e.g. image recognition \cite{farabetLearningHierarchicalFeatures2013, tompsonJointTrainingConvolutional2014, szegedyGoingDeeperConvolutions2015}, speech recognition \cite{krizhevskyImagenetClassificationDeep2012, mikolovStrategiesTrainingLarge2011, hintonDeepNeuralNetworks2012, sainathDeepConvolutionalNeural2013}
and Natural Language Processing \cite{collobertNaturalLanguageProcessing2011, bordesQuestionAnsweringSubgraph2014, jeanUsingVeryLarge2014, sutskeverSequenceSequenceLearning2014}. Much of this success can be accredited to the fact that Deep Learning does not require manual feature engineering, but is able to automatically extract discriminative features from raw data input \cite{najafabadiDeepLearningApplications2015}. The advantage of being able to apply algorithms on raw data and not being dependent on handcrafted features has led to studies investigating the effectiveness of Deep Learning in HAR, which e.g. suggested different architectures \cite{ordonezDeepConvolutionalLSTM2016, hammerlaPDDiseaseState2015, haMultiModalConvolutionalNeural2015, yangDeepConvolutionalNeural2015, xuInnoHARDeepNeural2019, chenLSTMNetworksMobile2016, yaoDeepSenseUnifiedDeep2017, ronaoHumanActivityRecognition2016, leeHumanActivityRecognition2017, haConvolutionalNeuralNetworks2016}, evaluated the generality of architectures  \cite{hammerlaDeepConvolutionalRecurrent2016} and assessed the applicability in real-world scenarios \cite{laneCanDeepLearning2015, hammerlaPDDiseaseState2015, jiangHumanActivityRecognition2015, guanEnsemblesDeepLSTM2017}.

\emph{DeepConvLSTM.}
One popular HAR Deep Learning architecture is the DeepConvLSTM which combines both convolutional and Long-Short-Term-Memory (LSTM) layers \cite{ordonezDeepConvolutionalLSTM2016}. By combining both types of layers the network is able to automatically extract discriminative features and model temporal dependencies. 
The idea of  Ordonez et al. \cite{ordonezDeepConvolutionalLSTM2016} to combine recurrent and convolutional layers served as the basis for several further improvements in subsequent works. 
For example, Murahari and Pl{\"o}tz \cite{murahariAttentionModelsHuman2018} improved results by appending attention layers to the original architecture. 
Xi et al. \cite{xiDeepDilatedConvolution2018} improved the performance of the original DeepConvLSTM architecture by adding dilated convolution layers in addition to normal convolution layers. Demonstrating a different way of combining recurrent and convolutional layers, Xu et al. \cite{xuInnoHARDeepNeural2019} proposed \textit{InnoHAR} which combines Inception modules based on GoogLeNet \cite{szegedyGoingDeeperConvolutions2015} and Gated Recurrent Unit (GRU) layers. Xia et al. \cite{xiaLSTMCNNArchitectureHuman2020} suggested to first apply a 2-layered LSTM whose output is then fed to two convolution layers and extended the architecture with a max pooling, global average pooling and batch normalization layer. Kim and Cho \cite{kimPredictingResidentialEnergy2019} make use of the architecture proposed by \cite{zhouCLSTMNeuralNetwork2015} which, similar to our approach, employs a one-layered LSTM. To our knowledge the work of Kim and Cho \cite{kimPredictingResidentialEnergy2019} is the only work which applied a one-layered LSTM within a variation of the DeepConvLSTM archtitecture.

\emph{LSTMs and sequential data.} 
Upon the experiments conducted by Karpathy et al. \cite{karpathyVisualizingUnderstandingRecurrent2015}, Chen et al. claim within their recent survey paper that ''the depth of an effective LSTM-based RNN needs to be at least two when processing sequential data'' \cite{chenDeepLearningSensorbased2021}. Within this paper we investigated the effect of employing a 1-layered instead of a 2-layered LSTM within the DeepConvLSTM architecture. As Karpathy et al. \cite{karpathyVisualizingUnderstandingRecurrent2015} obtained their results using character-level language models, i.e. text data, our paper aims at challenging the belief that their claim is applicable to sensor-based HAR.

\section{Methodology}

Contrary to the belief that one needs at least a two-layered LSTM when dealing with sequential data \cite{karpathyVisualizingUnderstandingRecurrent2015}, we propose to change the DeepConvLSTM to have a one-layered LSTM. Figure \ref{fig:architecture} illustrates the suggested architecture change. We further investigate the varying amount of hidden units in the LSTM layers during experiments.

\begin{figure*}[h]
  \centering
  \includegraphics[width=0.99\linewidth]{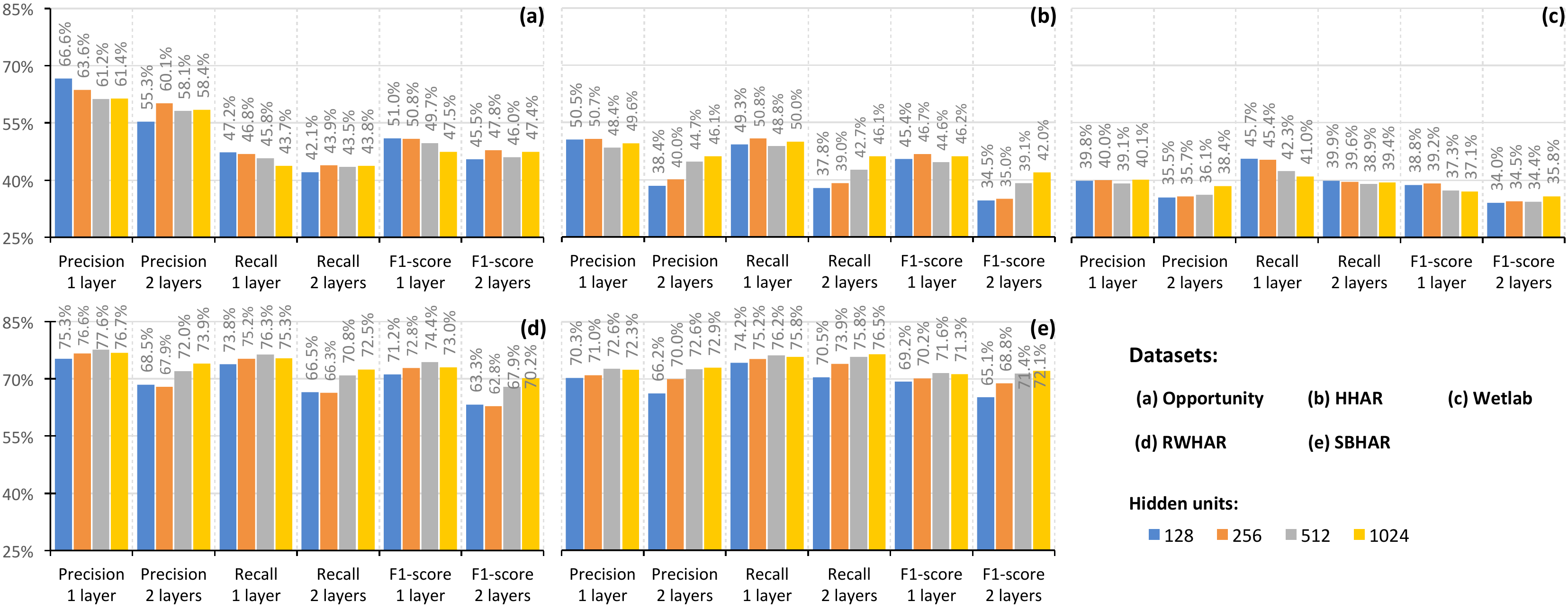}
  \caption{Results obtained on the (a) Opportunity, (b) HHAR, (c) Wetlab, (d) RWHAR and (e) SBHAR datasets using the 8 variations of the DeepConvLSTM \cite{ordonezDeepConvolutionalLSTM2016} architecture, grouped by 1- and 2-layered variations and color-coded by how many hidden units were employed in the LSTM layer(s). Results are given as the average precision, recall and F1-score across 5 runs using a set of 5 varying seeds. The 1-layered architecture variants continuously outperform the 2-layered ones for the HHAR, Wetlab and RWHAR dataset; The worst performing 1-layered variant even still outperforms the best performing 2-layered variant. For the Opportunity dataset, the 1-layered variant outperforms the 2-layered for 128, 256 and 512 hidden units and is on par for 1024 hidden units. For the SBHAR dataset, the 1-layered variant only outperforms for 128 and 256 hidden units and is on par with the 2-layered version when using 512 and 1024 hidden units.} 
  \label{fig:results}
\end{figure*}

\subsection{Datasets}
To validate our architecture change, we chose to use the preprocessed version of the Opportunity dataset \cite{roggenCollectingComplexActivity2010} as used by Ordonez et al. \cite{ordonezDeepConvolutionalLSTM2016}, as well as four popular HAR datasets, namely the Wetlab \cite{schollWearablesWetLab2015}, RealWorld HAR (RWHAR) \cite{sztylerOnBodyLocalizationWearable2016}, Smartphone-Based Recognition of Human Activities and Postural Transitions (SBHAR) \cite{reyes-ortizTransitionAwareHumanActivity2016} and the Heterogeneity Activity Recognition (HHAR) \cite{stisenSmartDevicesAre2015} dataset.

\emph{Opportunity.} The Opportunity datasets consists of 4 individuals performing a set of activities of daily living \cite{roggenCollectingComplexActivity2010}. For the gesture recognition challenge of the dataset, there are 18 classes which are to be predicted (\textit{open/ close door 1 and 2, fridge, dishwasher and drawer 1, 2 and 3, clean table, drink from cup and toggle switch}) as well as a \textit{null} class. During the experiments, we applied the same preprocessing as well as train-test split as suggested by Ordonez et al. \cite{ordonezDeepConvolutionalLSTM2016}. The resulting preprocessed dataset consists of in total 113 feature channels each representing an individual sensor axis from body-worn accelerometers and inertial measurement units (combining 3D accelerometers, gyroscopes and magnetometers).

\emph{Wetlab.} The Wetlab dataset consists of 22 participants performing two DNA extraction experiment within a wetlab environment \cite{schollWearablesWetLab2015}. During experiments subjects were equipped with a wrist-worn sensing unit capturing 3D acceleration data with a sampling rate of $50Hz$. Using recorded video footage, Scholl et al. identified 9 different actions (\textit{cutting, inverting, peeling, pestling, pipetting, pouring, stirring, transfer}) which, along the \textit{null} class, are the target labels to be predicted within this dataset.

\emph{RealWorld (HAR).} The RWHAR dataset contains data of 15 participants performing 8 different activities (\textit{walking upstairs, walking downstairs, jumping, lying, standing, sitting, running, walking}) as well as a \textit{null} class \cite{sztylerOnBodyLocalizationWearable2016}. In order to make results more comparable to the Wetlab dataset, we chose to only use 3D acceleration data captured by a wrist-worn sensor which samples the data at $50Hz$. 

\emph{SBHAR.} The SBHAR dataset consists of 30 participants performing activities of daily living (\textit{standing, sitting, lying, walking, walking downstairs, walking upstairs}) \cite{reyes-ortizTransitionAwareHumanActivity2016}. In addition to the 6 activities, class labels also include 6 postural transitions (\textit{stand-to-sit, sit-to-stand, sit-to-lie, lie-to-sit, stand-to-lie, lie-to-stand}) as well as a \textit{null} class. For the same reason as mentioned with the RWHAR dataset, we only used the raw 3D acceleration sensor data, which was sampled at $50Hz$. 

\emph{HHAR.} Similar to the RWHAR dataset, the HHAR dataset contains data of 9 human participants performing activities of daily living. There are 6 activities (\textit{biking, sitting, standing, walking, walking upstairs and downstairs}) and a \textit{null} class which are to be predicted \cite{stisenSmartDevicesAre2015}. As for the previous datasets, we only use the 3D-acceleration data obtained from a wrist-worn sensor unit, which was sampled at $100Hz$. 

\subsection{Training}
To justify our architecture and illustrate the effectiveness of said change to it, we compare two variations of the DeepConvLSTM architecture with each other: One equipped with a single-layer LSTM, and one equipped with a two-layer LSTM. We further vary the number of hidden units, more specifically using in both variations either 128, 256, 512 or 1024 hidden units per LSTM layer. This leaves us with a total of 8 variations that were evaluated, for all five datasets. It  is important to note that only the LSTM was altered. Specifications of other layers, such as the number of convolution layers ($4$), the number of convolution filters ($64$) and the dropout rate ($0.5$), were left unchanged from the original DeepConvLSTM architecture. 

To train our networks we employ a sliding window approach similar to the one as seen in \cite{ordonezDeepConvolutionalLSTM2016}. To obtain suitable set of hyperparameters we evaluated multiple settings based on results obtained from using the Wetlab dataset as input. We identified a sliding window of $1$ second with a $60\%$ overlap to be most suitable. Unlike Ordonez et al. \cite{ordonezDeepConvolutionalLSTM2016} we use the \textit{Adam} optimizer with a smaller weight decay ($1e^{-6}$) and learning rate ($1e^{-4}$) and initialize the network weights using the Glorot initialization \cite{glorotUnderstandingDifficultyTraining2010}. During training we computed the loss using a weighted cross-entropy loss to enable the networks to learn also imbalanced datasets (e.g. Wetlab). Since the focus of our paper lies on evaluating the change to the architecture, we did not perform any hypertuning on the RWHAR, SBHAR and HHAR dataset and thus kept hyperparameters consistent across datasets. Nevertheless, unlike the other 3 datasets, the HHAR was sampled at $100Hz$. We thus increased the convolutional filter size by a factor of two to be $21$, to maintain the relation between convolutional filter and sliding window size and capture the same amount of information with each filter across all datasets. We used the same set of hyperparameters for the Opportunity dataset as was used for the other datasets to allow a comparison across the different types of activity recognition scenarios, and changed the sliding window size and overlap to be identical with the one employed in \cite{ordonezDeepConvolutionalLSTM2016}, i.e. $0.5$ seconds with a $50\%$ overlap.

\subsection{Results}
For the Wetlab, RWHAR, HHAR, SBHAR and HHAR dataset, results were obtained using Leave-One-Subject-Out (LOSO) cross-validation. This means that each subject was being used as the validation set exactly one time while all other subjects were used as trainig data. The final validation results are then the average across all subjects. Using LOSO cross-validation ensures that obtained results were not a product of overfitting on subject-specific traits. As we are using the same preprocessed version of the Opportunity dataset as introduced in the original work \cite{ordonezDeepConvolutionalLSTM2016}, we also employed the same train/ test split and trained the networks using said inputs.

Figure \ref{fig:results} shows results for all five datasets using varying hidden units and either a 1- or 2-layered LSTM within the DeepConvLSTM architecture. Results were obtained by averaging across 5 runs with a set of 5 varying random seeds. We report standard evaluation metrics, namely precision, recall and F1-score.  

As we can see in Figure \ref{fig:results} architectures which employ a 1-layered LSTM overall outperforms architectures which employ a 2-layered LSTM across four (Opportunity, Wetlab, HHAR and SBHAR) out of five datasets. Only for the case of $1024$ employed hidden units and using the Opportunity dataset as input, our architecture is on par with the original DeepConvLSTM architecture. For the Wetlab, RWHAR and HHAR dataset we can even see that the best performing 2-layered LSTM architecture variation performs worse, as far as precision, recall and F1-score, than the worst performing 1-layered LSTM architecture variation. Only for the SBHAR dataset our architecture is marginally worse than the original one when using $1024$ hidden units within the LSTM layers. 

By removing the second LSTM layer we are able to get performance increases up to $11.7\%$ in F1-score (HHAR, 256 hidden units). Looking at Figure \ref{fig:results} one can see that the performance increase is the largest for variations employing $128$ and $256$ units and steadily decreases for larger amounts of hidden units. While one can see that the performance difference was not as significant for the SBHAR dataset when employing $512$ hidden units and even slightly negative for the $1024$ hidden units architecture, we nevertheless argue that both architecture variations are at least on par due to the fact that one has to consider statistical variance. Over the five runs, we witness an average standard deviation of $1.67\%$ ($1.53\%$) for precision, $1.69\%$ ($1.44\%$) for recall and $1.65\%$ ($1.38\%$) for the F1-score, using the 1-layered (2-layered) version of the architecture, respectively. The exact averages per architecture variation can be found in the repository (see \url{https://github.com/mariusbock/dl-for-har}).

As HAR datasets are relatively small in size compared to other popular Deep Learning datasets, e.g. \cite{dengImageNetLargeScaleHierarchical2009}, larger networks are more likely to be prone to overfitting. Though we witnessed this trend for the HHAR and Opportunity dataset, by monitoring the generalization gap (i.e. the difference in loss, accuracy, precision, recall and F1-score between the train and validation set), we did not see any indication that the 2-layered architecture variations are more likely to overfit and saw both architectures equally suffering from this phenomenon.


With the suggested removal of one LSTM layer we are decreasing the complexity of the DeepConvLSTM architecture. 
Assuming both layers in the original LSTM have the same number of hidden units $h$, the number of learnable parameters $p_{2}$ within the LSTM of DeepConvLSTM is 
\begin{math}
p_{2} = 4 s h + 8h + 12h^2
\end{math}
with $s$ denoting the size of the sliding window. 
By removing the second layer, the number of LSTM parameters in our architecture reduces to 
\begin{math}
  p_{1} = 4 s h + 4h + 4 h^2,
\end{math}
which shows that the increase in parameters with an increasing number of hidden units is dominated by $4h^2$ for $p_1$ opposed to $12 h^2$ for $p_2$.
In our experiments the removal of one LSTM layer roughly equated in $63\%$ fewer learnable LSTM parameters across all experiments. This decrease in complexity can be also seen in decreased runtimes using our architecture compared to the original DeepConvLSTM architecture. On average we are seeing a decrease in runtime of $30\%$ across all experiments with the difference increasing (going up to as much as $48\%$) the more hidden units are employed and the larger the input dataset is.

\section{Conclusion and Future Directions}

We proposed to change the DeepConvLSTM architecture as frequently used in activity recognition to employ a 1-layered instead of a 2-layered LSTM. We validated our architecture change with experiments using 5 publicly available datasets \cite{roggenCollectingComplexActivity2010, schollWearablesWetLab2015, sztylerOnBodyLocalizationWearable2016, reyes-ortizTransitionAwareHumanActivity2016, stisenSmartDevicesAre2015} with varying numbers of hidden units within the LSTM.

Results show that for 4 out of 5 datasets, one LSTM layer consistently outperforms the original DeepConvLSTM architecture \cite{ordonezDeepConvolutionalLSTM2016} in terms of precision, recall and F1-score. For \cite{reyes-ortizTransitionAwareHumanActivity2016}, it outperforms the original 2-layer architecture only for LSTMs with smaller number of hidden units, and is on par for larger LSTMs.
With our suggested removal of the second LSTM layer, we are able to decrease the number of learnable parameters of the LSTM within the DeepConvLSTM \cite{ordonezDeepConvolutionalLSTM2016} by an average factor of $62\%$ and are able to decrease training time on average by a factor of $38\%$, going up to as much as $48\%$ for larger networks (with more hidden units). 

These findings contradict with the belief that one needs at least a 2-layered LSTM when dealing with sequential data \cite{karpathyVisualizingUnderstandingRecurrent2015}. As this belief finds its origins in the area of text analytics, we suggest that it might not be applicable to sensor-based HAR.
Our choice of datasets to use for our experiments was driven by testing our hypotheses against datasets containing complex (Wetlab and Opportunity), transitional (SBHAR) and simpler/ periodical activities (SBHAR, HHAR and RWHAR). Looking at the results one can see that the hypothesis holds for all three types of activities. Furthermore, with the Opportunity dataset and its 113 feature dimensions we showed that a performance increase can be also witnessed when using more than one sensor. Therefore we currently cannot clearly identify specific reasons for why 1-layered LSTMs are performing better within the setting of HAR. 
Next steps within this research include investigating whether the hypothesis holds true for larger HAR datasets to further rule our overfitting being a reason for the perfomance increase. Using e.g. a feature extraction tool as seen in \cite{kwonIMUTubeAutomaticExtraction2020a} could be used to create such larger datasets. Further, we will analyse how 1-layered LSTMs learn compared to 2-layered ones within the setting of HAR by replicating a similar analysis as performed by Karparthy et al. \cite{karpathyVisualizingUnderstandingRecurrent2015}.

This paper's architecture and experiment scripts are publicly downloadable via \url{https://github.com/mariusbock/dl-for-har}, to support the continuation, replication, as well as further analysis of our architecture and experiments.

\bibliographystyle{ACM-Reference-Format}
\bibliography{main}

\end{document}